# Open F-branes

Machiko Hatsuda

*Department of Radiological Technology, Faculty of Health Science, Juntendo University*
*Yushima, Bunkyou-ku, Tokyo 113-0034, Japan*
*KEK Theory Center, High Energy Accelerator Research Organization*
*Tsukuba, Ibaraki 305-0801, Japan*

and

Warren Siegel

*CNYITP*
*Stony Brook University*

March 1, 2022

**Abstract**

We include in F-theory, through open Type I F-theory branes (F-branes), string theories with N = 1 supersymmetry, both Type I and heterotic. Type I branes are distinguished from Type II by worldvolume parity projection. The same open Type I branes describe both open Type I superstrings and closed heterotic upon different sectionings from F-branes to worldsheets, while closed Type I superstrings arise from closed Type I branes. (Type II superstrings come from closed Type II branes, as described previously.)

F-theory manifests the exceptional-group U-duality symmetry, with all massless bosonic fields in a single gauge coset. This coset branches to the usual bosonic supergravity fields upon sectioning. We examine in detail the simple case of D = 3 F-theory: Parity projection reduces the Type II coset SL(5)/SO(3,2) to the Type I coset SO(3,3)/SO(2,1)$^2$ = SL(4)/SO(2,2).

# Contents

# 1 Type II F-theory

## 1.1 Couplings and massless fields

In T-duality, (a component of) the spacetime metric is transformed into its inverse (up to $B$-field dependence). This implies the usual R → 1/R transformation, as the metric of the compactified dimensions can be expressed in terms of "angles" as $ds^2 = R^2 d\theta^2$. More properly, the gravitational constant (in Einstein frame, or $\alpha'$ in string frame) can always be absorbed into the metric, reappearing through its vacuum value. Thus R and $\alpha'$ appear in the combination $R^2/\alpha'$, so actually $R^2/\alpha' \to \alpha'/R^2$ (and R and $\alpha'$ always appear in this combination in AdS$_5\times$S$^5$).

In S-duality (in Type IIB), the dilaton appears as part of a metric of scalars, which transforms into its inverse in the same way. (Only the coset is different.) Now it's the string coupling g that appears as the vacuum value. Thus $\alpha'$ and g, and their dualities, appear in a completely analogous way in the field theory action, but differently in the first-quantized string action. Also, in compactifying 11D supergravity to 10D Type IIA, the dilaton, and thus g, arises from a component of the metric associated with the compactification radius R.

In D<10 (after field-theory duality transformations), all the compactification scalars merge into a larger coset space (U-duality). The purpose of F-theory is to exhibit this (spontaneously broken) symmetry before compactification. The metric, scalars, and other massless bosonic fields then appear as parts of a higher-dimensional gauge field [1, 2] and their field theoretical formulation is the exceptional field theory [3–7]. As these symmetries relate the various superstring theories, F-theory is meant to unify them perturbatively.

## 1.2 Sectioning

The limit R → ∞ (or 0) makes winding modes (or momenta) infinitely massive, for each dimension of toroidal compactification. The Virasoro ($\mathcal{S}$) section condition is the remnant of the $L_0 = \bar{L}_0$ constraint that enforces the decompactification limit in any combination of R's going to 0 or ∞. This constraint appears already in T-theory, the manifestly T-duality covariant formulation of string theory [8]. Solving the constraint chooses the vacuum, producing either Type IIA or B superstring theory via spontaneous breakdown. There is also an intermediate Virasoro condition, linear in both zero and nonzero modes, that acts as a first-order form of the selfduality condition on the field strength of $X$.

The generalization [1, 2] appears in F-theory, as a vector in the "spatial" directions of the worldvolume [9]. Thus string theories related by S-duality (different vacua for Type IIB) are also chosen. F-theory also has a Gauss ($\mathcal{U}$) section condition [9] that mutually constrains spacetime and the worldvolume. It's the zero-mode part of Gauss's law for the gauge field $X$.

# 2  Type I F-theory

## 2.1  Geometry

As for strings, we'll classify branes as Types I and II, corresponding to the number of supersymmetries:

(I) Like Type I strings, the states of Type I branes, both open and closed, are constrained to be (graded) symmetric under $\sigma_1$ reflection ("parity"). In particular, this is true for open branes because of boundary conditions. These open branes describe both heterotic and open Type I strings, while the closed branes describe closed Type I strings.

(II) Type II branes are only closed, and have no parity projection. As described in previous papers, these describe Type IIA and B strings (as well as M-theory) [9–17].

For simplicity, we consider closed branes to have the topology (for fixed $\tau$) of a torus, and open branes to have the topology of a torus of 1 less dimension × a line segment. (The line segment $\sigma_1$ is sometimes described as the simple orbifold $S^1/Z_2$.) Thus we discuss branes with only 0 (closed) or 2 (open) boundaries. "Internal symmetry" degrees of freedom live on the boundaries; in general these contribute to the worldvolume energy-momentum tensor (generalized Virasoro operators).

Our description of heterotic strings in terms of open branes will follow that of Hořava and Witten [18], with a few important differences:

(1) Our branes are fundamental, not solitons. This should allow perturbative S-matrix calculations in the higher dimensions.

(2) Upon double dimensional reduction, the Lorentz symmetry of spacetime was found tied to that of the worldvolume (as applied to $\Theta$); for us the 2 are the same before compactification. This is related to the fact that STU symmetries are manifest without compactification, by construction.

(3) Instead of orbifolding spacetime ($S^1/Z_2$), which in turn orbifolds the embedded worldvolume, we orbifold only the worldvolume. This has several advantages:

   (a) The F-branes can end on space-filling branes, not just 9-branes. So ordinary Type I superstrings can be obtained, not just Type IA (I′). Thus both $E_8 \times E_8$ and SO(32) are allowed. (But this does not make obvious why Type I $E_8 \times E_8$ is not allowed.)

   (b) We derive Type I and heterotic strings by worldvolume compactification, not (explicitly) spacetime (although the latter may be implied by constraints). However, since the worldvolume ($\sigma$) derivatives represent the same group as the spacetime ($X$) derivatives, their covariantizations both have vielbeins that are elements of the coset space containing the bosonic spacetime fields [9]. Thus the effects of both worldvolume reduction and the parity symmetry that projects Type II to Type I F-branes can be seen directly on the bosonic fields, without the need to examine spacetime compactification.

(c) The worldvolume of the closed F-brane ("donut") can't be compactified along its long(est) circle (without double compactification), unlike the open F-brane ("cylinder"). So open F-branes can yield either Type I open or heterotic (closed), while closed F-branes can give only Type I closed:

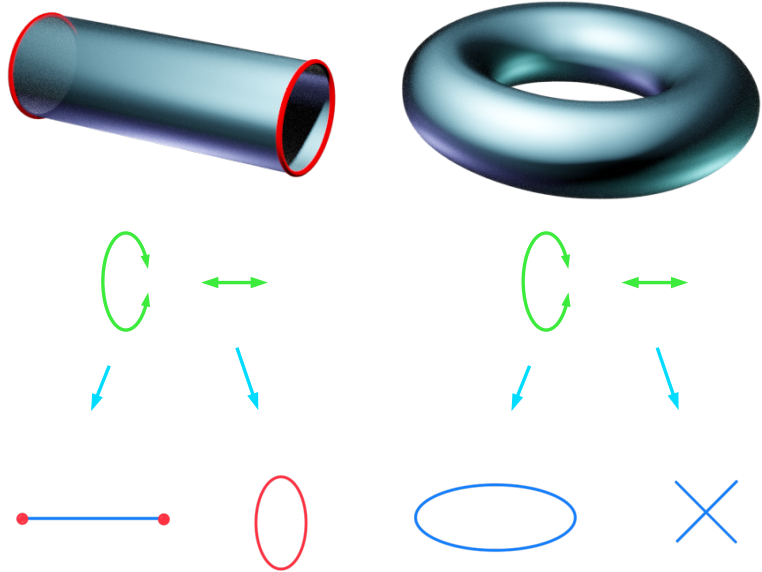

Type I open ($\sigma_1$)   heterotic ($\sigma_2$)   Type I closed ($\sigma_1$)   *(not)*

Internal symmetry regions are in red. We have illustrated only 2 of the worldvolume dimensions, $\sigma_1$ and $\sigma_2$ (and not $\tau$ nor the other $\sigma$'s).

For the open F-brane, $\sigma_1$ is the coordinate for the line segment, and $\sigma_2$ is one of the toroidal coordinates. For the closed F-brane, $\sigma_1$ is for the major circumference, and $\sigma_2$ is for the minor one, because closed branes can form by joining ends of open ones.

Heterotic and open Type I strings follow from open branes by 2 different choices of section: If $\sigma_2$ is killed by sectioning (along with the rest of the toroidal coordinates), we get an open string, and each boundary is compressed to a point. On the other hand, if $\sigma_1$ is eliminated (along with the extra toroidal coordinates), we get a closed string, and the 2 boundaries are compressed into the bulk of the string, together.

Gravity arising from open F-branes might appear unusual, but a similar situation occurs in string theory, where the open-closed string 2-point function implies that closed string fields (in particular, the metric) already are contained in the open string field (the metric from classically massive spin-2 fields) [19].

To treat SO(32) vs. $E_8 \times E_8$ by sectioning, we can choose the boundaries of open F-branes to live in 496 extra dimensions. The (boundary) background metric and 2-form for this space can be chosen to give the current algebras for these 2 groups with appropriate level number, in a way that manifests these symmetries, but spontaneously breaks SO(496) [20].

## 2.2 Parity

For (super)strings, Type II is reduced to Type I by constraining *states* (or background fields, not just the action) to be invariant under worldsheet parity, $\sigma \to -\sigma$. This is sometimes called a type of orbifolding, but is more a type of projection, like GSO (which is for the worldsheet symmetry of $2\pi$ rotation, which affects only fermions). In T-theory, this is accompanied by switching the independent $X$'s for left and right modes, to preserve their (anti)selfduality condition. The selfdual and anti-selfdual currents are $\partial_\tau X^N \hat{\eta}_{NM} \pm \partial_\sigma X^N \eta_{NM}$ where $\hat{\eta}_{MN}$ and $\eta_{MN}$ are the double Minkowski metric and the O(D,D) invariant metric respectively. The selfduality condition is that the anti-selfdual current is 0. It relates the dual coordinate $X_m$ and the usual coordinate $X^m$ in the co/contravariant basis $M = (_m, {}^m)$. The selfduality condition is preserved under the worldsheet parity transformation $\sigma \to -\sigma$ together with interchanging the left and right coordinates in the left/right basis.

In F-theory both the target space coordinate $X$ and the worldvolume coordinates $\sigma$ are different representations of the exceptional symmetry groups, and currents are field strengths $\partial_\sigma X$. A naive parity transformation would violate the selfduality constraint on the $X$ field strengths. (For spacetime dimension D > 1, worldvolume dimension d is always even.) So instead we need a rotation as an even number of reflections. For consistency with Type I and heterotic strings, this should change the sign of $\sigma_1$ but not $\tau$ nor $\sigma_2$, so it's most symmetry-preserving to change signs for *all* of $\sigma$ *except* $\tau$ and $\sigma_2$. This rotation then implies the same rotation on $X$.

For example, for D = 3 (d = 6), where in the Lagrangian formalism we have $\sigma_m$ and $X^{mn}$ (antisymmetric), $\mathcal{P}$(arity) is represented on the indices $m, n$ by the sign changes

$$\mathcal{P}: \quad (0, 2) \to +, \quad (1, 3, 4, 5) \to - \tag{2.1}$$

as a rotation in the Lagrangian symmetry L = SO(3,3). This breaks that symmetry to SO(2,2).

## 2.3 Symmetries

This parity (as well as boundary conditions), since it distinguishes only $\sigma_2$ (as well as $\tau$), effectively reduces symmetries of the Hamiltonian by 1 worldvolume dimension, $\Delta d = 1$. Generally $\sigma$, including $\tau$, is a bispinor: $\sigma^{(\alpha\beta)}$, $\sigma^\alpha{}_{\bar{\beta}}$, or $\sigma^{[\alpha\beta]}$ [13]:

|              | H   | SO | SL | Sp |              | H'(SU) |
|--------------|-----|----|----|-----|--------------|--------|
| Real         |     | 1  | 2  | 3  | $X^{(\alpha\beta)}$ | SO     |
| Complex      |     | 0  |    | 4  | $X^{\alpha\dot\beta}$ | SU     |
| Pseudo-real  |     | 7  | 6  | 5  | $X^{[\alpha\beta]}$ | USp    |

$$\sigma^{(\alpha\beta)} \quad \sigma^{\alpha}{}_{\bar\beta} \quad \sigma^{[\alpha\beta]}$$
$$Y^{(\alpha'\beta')} \quad Y^{\alpha'}{}_{\bar\beta'} \quad Y^{[\alpha'\beta']}$$

Since fermions are representation of the tangent space symmetry H⊂G, bosonic coordinates are bispinors of H in the superalgebra. Numbers are spacetime dimensions of corresponding string theories. The primed objects are for extended supersymmetry for that dimension, with additional D'-dimensional spacetime coordinates Y. The (spacetime) spinor size increases by a factor of 2 from one oval to the next. In the oval for 3 and 4 X's are real and complex representations for D=3 and 4 theories respectively. As the exceptional group G-symmetry in the Hamiltonian formalism is enlarged to F-symmetry in the Lagrangian formalism, its tangent space symmetry H is also enlarged to L-symmetry. Going from the Lagrangian to the Hamiltonian formalism breaks L-symmetry (always a GL group) to H by picking the matrix coefficient of $\tau$ in the matrix $\sigma$ to be a group metric for $\mathcal{G}$ = SO, SL, or Sp. (See also Appendix A.) Doing the same for $\sigma_2$ yields 2 of the same type of group metric, reducing the H group $\mathcal{G}(2n)$ to $H_I = \mathcal{G}(n)^2 = \mathcal{G}_+ \times \mathcal{G}_-$, where $\mathcal{G}(n)$ is the usual covering group of SO(D−1,1), and ± will eventually have the usual string identifications of left and right.

For Type II theories F-theory, which has the most symmetry, is broken to either M-theory (which has (D+1)-dimensional N = 1 supergravity as its low-energy limit) or T-theory (with manifest T-duality) by solving either an $\mathcal{S}$ or $\mathcal{U}$ sectioning condition (reducing both $X$ and $\sigma$), or S-theory (the standard formulation of superstring theory) by solving both.

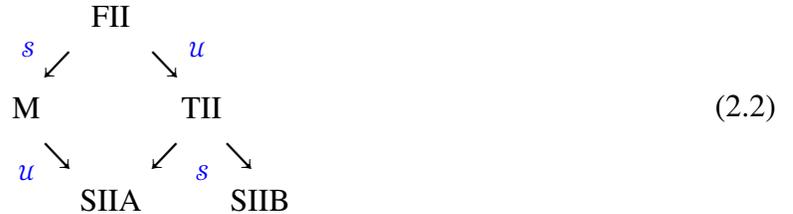

$$\begin{array}{ccc} & \text{FII} & \\ {}^{\mathcal{S}}\swarrow & & \searrow^{\mathcal{U}} \\ \text{M} & & \text{TII} \\ {}_{\mathcal{U}}\searrow & \swarrow_{\mathcal{S}} & \searrow \\ & \text{SIIA} \quad \text{SIIB} & \end{array} \qquad (2.2)$$

Since only $\mathcal{U}$ sectioning forces reduction of $\sigma$, the brane pictures appearing above (cylinder and donut) apply to both F-theory and M-theory, while the string pictures (worldsheet only) apply to both T-theory and S-theory.

F-symmetry has the Dynkin diagram [14]:

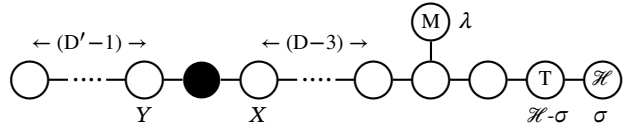

D+D′ = 10 is the maximal case, which has 12 nodes. The black node is dropped unless D′ = 0. Remove node M or T for M or T-theory, both for S-theory. Remove $\mathcal{H}$ for the Hamiltonian subgroup of any of these Lagrangian groups. $Y, X, \lambda$ (parameters for $X$'s gauge invariance), $\sigma$, and $\mathcal{H}$-$\sigma$ are the nodes for their representations. The subgroup L of the F/L coset is a Wick rotation of the maximal compact subgroup of F. (It also corresponds to the covering group of the Lorentz group SO(D−1,1), but with twice the argument, due to doubled spinor indices on supersymmetry. Embedding the SO(D,D) of Type II T-theory in $E_{D+1}$ was also considered in [21], but there $E_{D+1}$ was considered as a symmetry of the superstring supergravity of D+1, not D, dimensions.)

The effect of going from the Lagrangian formalism to the Hamiltonian is to eliminate $\tau$, and thus reduce the Lagrangian symmetry F to the Hamiltonian symmetry G, the symmetry of the background fields, which removes the last node on the right. Projecting from Type II to Type I F-branes then distinguishes $\sigma_2$ (parity changes signs for all worldvolume coordinates except $\tau$ and $\sigma_2$), knocking off the next node on the right, which breaks the symmetry down to $G_{FI}$ = SO(D,D). Since the heterotic string contracts all but $\tau$ and $\sigma_2$, it maintains that symmetry. But Type I strings contract $\sigma_2$ instead of $\sigma_1$, killing the next node on the right, further breaking the symmetry to $G_I$ = GL(D). Note that these reductions in $\sigma$ (d) drop nodes from the right, whereas reduction in $X$ (D) drops nodes from the left.

The resulting string of cosets, from consecutively eliminating $\tau, \sigma_1, \sigma_2$ from the symmetry, is

$$\frac{F}{L} \quad \rightarrow \quad \frac{G}{H} \quad \rightarrow \quad \frac{G_{FI}}{H_{FI}} = \frac{SO(D,D)}{SO(D-1,1)^2} \quad \rightarrow \quad \frac{G_I}{H_I} = \frac{GL(D)}{SO(D-1,1)} \qquad (2.3)$$

(For each of these cosets A/B, B is a Wick rotation of the maximal compact subgroup of the split group A.)

It might seem strange that $G_{FI}/H_{FI}$ is also G/H for Type II T-theory (both A and B), until one remembers that the complete bosonic sector of N = 1 stringy supergravity is the same as the NS-NS sector of N = 2, as easily seen from the heterotic construction: vector × vector, no symmetrization. (If D ≠ 3,4,6,10, there would also be scalars × vector, since supersymmetry requires vector multiplet × vector.) The coset $G_I/H_I$ is then the coset of the NS-NS sector of Type I T-theory, namely just gravity + physical scalar (as the determinant of the metric).

In the string frame for T-theory, the "dilaton" is an unphysical, compensating scalar from ghost × ghost. In this reduction, applying parity to G/H restricts it to not just $G_{FI}/H_{FI}$, but also the dilaton, while not eliminating spacetime coordinates. Then, applying sectioning reduces the spacetime coordinates, but for Type I simply breaks that coset into the metric (the last coset) + the 2-form (but doesn't break the coset for heterotic).

$$\frac{G}{H} \rightarrow GL(1)\frac{SO(D,D)}{SO(D-1,1)^2} = \text{dilaton} \oplus \frac{GL(D)}{SO(D-1,1)} \oplus \text{2-form} \quad (2.4)$$

Although the fields and number of coordinates are the same for Type I and heterotic T-theories, the symmetries are different because different coordinates have been eliminated in the sectioning.

Cosets of bosonic fields:

(i) G/H: Type II F-theory

(ii) SO(D,D)/SO(D−1,1)$^2$: Type I F-theory, heterotic supergravity, NS-NS Type II T-theory

(iii) GL(D)/SO(D−1,1): NS-NS Type I T-theory

For example:

$$D = 3: \quad \frac{SL(6)}{SO(3,3)} \rightarrow \frac{SL(5)}{SO(3,2)} \rightarrow \frac{SL(4)}{SO(2,2)} = \frac{SO(3,3)}{SO(2,1)^2} \rightarrow \frac{SL(3)}{SO(2,1)} \quad (2.5)$$

$$D = 4: \quad \frac{SO(6,6)}{SO(6,C)} \rightarrow \frac{SO(5,5)}{SO(5,C)} \rightarrow \frac{SO(4,4)}{SO(4,C)} = \frac{SO(4,4)}{SL(2,C)^2} = \frac{SO(4,4)}{SO(3,1)^2} \rightarrow \frac{SL(4)}{SO(3,1)} \quad (2.6)$$

Couplings can be studied through these reductions, as vacuum values of compactification scalars. Since the same coset fields appear as metrics for both the worldvolume and spacetime, these scalars measure distances in both.

## 3 M-theory

Although we will concentrate on T-theory because it incorporates the greater number of S-theories, we briefly divert to M-theory to see how the original Hořava-Witten treatment fits in. M-theory is best described [16] by a worldvolume theory where $X^M$ is a vector gauge field, so it carries the same index as $\sigma_M$. (We can also include a dual $X$, related by a duality constraint, useful for describing a background 6-form field.) Note that our arguments for M-theory and its relation to T-theory can be applied all the way to the critical dimension. (Because of the reliance on exceptional groups, detailed discussions for F-theory need to be specialized for each D, and in this paper we concentrate on D = 3 for simplicity.)

The bosonic fields are then the metric $G_{MN}$ and 3-form $A_{MNP}$, where the indices are for GL(D+1). (We ignore the 6-form, which can be relevant in higher D: See the counting arguments in Appendix B.) Parity projection is easy to apply, as fixing $\sigma_2$ fixes the corresponding index on the fields. (Background fields don't carry the $\tau$ index.) Separating $M = (2, m)$ for the GL(D) index $m$, we're left with fields with even numbers of GL(D) indices:

$$G_{mn}, \quad G_{22} = \varphi, \quad A_{2mn} = B_{mn} \tag{3.1}$$

We can call this theory M′-theory. It has the same $\sigma$'s and $X$'s as M-theory, but parity projection has reduced the number of fields, and consequently reduced the coset space from GL(D+1)/SO(D,1) to GL(D)/SO(D−1,1). Thus, although still a (D+1)-dimensional theory, it has the symmetries of D-dimensional N = 1 supergravity. This is analogous to the situation described by Hořava and Witten, although our worldvolume is larger, and $X$-parity projection is a consequence of $\sigma$-parity projection.

However, according to our picture, M′-theory can also be obtained by first applying parity projection, and then $\mathcal{S}$ sectioning:

$$\begin{array}{ccc} & \text{FII} & \\ \mathcal{P} \swarrow & & \searrow \mathcal{S} \\ \text{FI} & & \text{M} \\ \searrow & & \swarrow \\ & \text{M}' & \end{array} \tag{3.2}$$

We next reduce to S-theory by applying $\mathcal{U}$ sectioning, the single constraint $\partial^M P_M = 0$. To obtain a D-dimensional theory, this requires

$$\partial^M P_M = 0 \quad \Rightarrow \quad \partial^M = \partial^2, \quad P_M = P_m \tag{3.3}$$

Thus the worldvolume is reduced to the worldsheet, while spacetime us reduced to D dimensions, while the above massless background fields are retained. According to our picture for parity preserving $(\tau, \sigma_2)$ while contracting all the rest, this is (the supergravity sector of) the heterotic string.

Other routes to heterotic S-theory will be described below. (N = 1 supersymmetric S-theories follow from different orderings of application of $\mathcal{P}, \mathcal{S}, \mathcal{U}$.)

# 4 T-theory

## 4.1 Type II

Before deriving Type I gravity (NS-NS sector) by $\mathcal{P}$ projection from Type II in T-theory, we first review Type II [8, 22, 23]. Working in the Hamiltonian formalism, we express the curved-space currents in terms of the flat-space currents as

$$\triangleright_A = E_A{}^M \mathring{\triangleright}_M, \quad \mathring{E}_A{}^M = \frac{1}{\sqrt{2}} \begin{pmatrix} \delta_a^m & -\eta_{am} \\ \delta_{\bar{a}}^m & \eta_{\bar{a}m} \end{pmatrix} \tag{4.1}$$

where the flat index $A = (a, \bar{a})$ is in a left/right basis, while the curved index $M = (_m, {}^m)$ is in a co/contra-variant basis. ($\mathring{E}$ is the matrix that converts between the 2 bases. The flat currents can be taken as $\mathring{\triangleright}_M = (P_m, X'^m)$ in a basis where $X$ is not doubled.) The SO(D,D) metric in these 2 bases is

$$\eta_{AB} = \begin{pmatrix} \eta_{ab} & 0 \\ 0 & -\eta_{\bar{a}\bar{b}} \end{pmatrix}, \quad \eta_{MN} = \begin{pmatrix} 0 & \delta_m^n \\ \delta_n^m & 0 \end{pmatrix} \tag{4.2}$$

Since the concept of left and right will not survive in F-theory, we'll generally work instead in a co/contra-variant basis also for flat indices, $A = (_a, {}^a)$. Then the metric is the same for both flat and curved indices, and the flat vielbein is simply the identity, but the local SO(D−1,1)² transformations are more complicated (not block diagonal).

The orthogonality condition on $E$ is

$$\eta^{AB} E_B{}^N \eta_{NM} = E_M{}^A \tag{4.3}$$

in terms of its inverse. In a gauge that breaks SO(D−1,1)² to the diagonal SO(D−1,1), a convenient form for the solution to orthogonality for the section $\partial^m = 0$, where $e$ and $B$ have the usual gauge transformations, is

$$E_A{}^M = \hat{E}_A{}^N B_N{}^M = \begin{pmatrix} e_a{}^n & 0 \\ 0 & e_n{}^a \end{pmatrix} \begin{pmatrix} \delta_n^m & B_{nm} \\ 0 & \delta_m^n \end{pmatrix} = \begin{pmatrix} e_a{}^m & e_a{}^n B_{nm} \\ 0 & e_m{}^a \end{pmatrix} \tag{4.4}$$

or for $\partial_m = 0$ (with the T-dual gauge transformations),

$$E_A{}^M = \begin{pmatrix} e_a{}^n & 0 \\ 0 & e_n{}^a \end{pmatrix} \begin{pmatrix} \delta_n^m & 0 \\ B^{nm} & \delta_m^n \end{pmatrix} = \begin{pmatrix} e_a{}^m & 0 \\ -B^{mn} e_n{}^a & e_m{}^a \end{pmatrix} \tag{4.5}$$

To treat more general sections, or describe both at the same time, we'll sometimes write this as

$$E_A{}^M = \begin{pmatrix} e_a{}^n & 0 \\ 0 & e_n{}^a \end{pmatrix} \begin{pmatrix} \delta_n^m & B_{nm} \\ B^{nm} & \delta_m^n \end{pmatrix} = \begin{pmatrix} e_a{}^m & e_a{}^n B_{nm} \\ -B^{mn} e_n{}^a & e_m{}^a \end{pmatrix}, \quad B^{mn} B_{np} = 0 \tag{4.6}$$

(maintaining orthogonality).

In the absence of fermions, one can also work directly in terms of the metric

$$M^{MN} = \hat{\eta}^{AB} E_A{}^M E_B{}^N, \quad \hat{\eta}_{AB} = \begin{pmatrix} \eta_{ab} & 0 \\ 0 & \eta_{\bar{a}\bar{b}} \end{pmatrix} \tag{4.7}$$

This produces the usual

$$M^{MN} = \begin{pmatrix} G^{mn} & G^{mp}B_{pn} \\ -B_{mp}G^{pn} & G_{mn} - B_{mp}G^{pq}B_{qn} \end{pmatrix} \tag{4.8}$$

or (as follows from T-duality $\partial^m \leftrightarrow \partial_m$)

$$M^{MN} = \begin{pmatrix} G^{mn} - B^{mp}G_{pq}B^{qn} & -B^{mp}G_{pn} \\ G_{mp}B^{pn} & G_{mn} \end{pmatrix} \tag{4.9}$$

and in the general case

$$M^{MN} = \begin{pmatrix} G^{mn} - B^{mp}G_{pq}B^{qn} & G^{mp}B_{pn} - B^{mp}G_{pn} \\ G_{mp}B^{pn} - B_{mp}G^{pn} & G_{mn} - B_{mp}G^{pq}B_{qn} \end{pmatrix} \tag{4.10}$$

The gauge transformations are

$$\delta E_A{}^M = \lambda^N \partial_N E_A{}^M - E_A{}^N (\partial_N \lambda^M - \partial^M \lambda_N) \tag{4.11}$$

with indices raised and lowered with $\eta$. The non-transport term is an orthogonal transformation, and so preserves orthogonality.

We then identify the scalar density dilaton as transforming so $\int \Phi^2$ is invariant,

$$\delta \Phi = \lambda^M \partial_M \Phi + \tfrac{1}{2} \Phi \partial_M \lambda^M \tag{4.12}$$

After sectioning, the density $\Phi$ relates to a scalar $\varphi$ as

$$\Phi^2 = \varphi^2 \sqrt{-G} \tag{4.13}$$

Besides the NS-NS sector, described by the coset SO(D,D)/SO(D−1,1)$^2$, and the ghost-ghost sector described by the dilaton, the bosons also include the R-R sector, described by an SO(D,D) Weyl spinor [21, 24–29]. It has its own gauge transformation and field strength

$$\delta R = \gamma^M \partial_M \Lambda, \quad F = \gamma^M \partial_M R \tag{4.14}$$

where the SO(D,D) $\gamma$-matrices satisfy

$$\{\gamma^M, \gamma^N\} = 2\eta^{MN} \tag{4.15}$$

(since $(\gamma^M \partial_M)^2 = \Box = 0$ is the section constraint). Its coordinate transformation can then be written gauge covariantly (a simple way to derive the noncovariant form) as

$$\delta R = \tfrac{1}{2}(\gamma^M \lambda_M) F = (\delta_0 + \delta_1) R \tag{4.16}$$

where

$$\delta_0 R = \lambda^M \partial_M R + \tfrac{1}{4}\gamma^{[M}\gamma^{N]}(\partial_M \lambda_N)R + \tfrac{1}{2}(\partial_M \lambda^M)R \,, \quad \delta_1 R = -\tfrac{1}{2}\gamma^M \partial_M(\gamma^N \lambda_N R) \qquad (4.17)$$

with a usual tensor density transformation $\delta_0$ plus a gauge transformation piece $\delta_1$. Since the R-R fields are weight $\tfrac{1}{2}$, their (quadratic) action is already weight 1 without coupling to the dilaton $\Phi$ (also weight $\tfrac{1}{2}$).

This spinor gauge field resembles the *coordinates X* of 4D F-theory [10]. On sectioning, the resulting formalism for bispinors resembles that of the so-called "Dirac-Kähler" formalism [30–33].

## 4.2  3D

For the case D = 3, we can use "spinor notation", and replace $A$ (and $M$) of SO(3,3) ($A$ = 1-6) with $[AB]$ of SL(4) ($A$ = 1-4), to avoid introducing too many new kinds of indices. In this notation $\gamma$-matrices, the flat-space metric, and the Levi-Civita tensor of SO(3,3) are all expressed in terms of Kronecker $\delta$'s and the $\varepsilon$-tensor of SL(4):

$$\gamma^{[AB]}_{CD} = \delta^{[A}_C \delta^{B]}_D \,, \quad \gamma^{[AB]CD} = \varepsilon^{ABCD} \qquad (4.18a)$$

$$\eta^{[AB][CD]} = \varepsilon^{ABCD} \,, \quad \varepsilon^{[AB][CD][EF][GH]} = \varepsilon^{[A[C[E[G}\varepsilon^{H]F]D]B]} \qquad (4.18b)$$

Then for the coset of NS-NS fields it's useful to work with the 4-representation $\mathcal{E}$ of SL(4) instead of the 6-representation $E$:

$$E_{AB}{}^{MN} = \mathcal{E}_{[A}{}^M \mathcal{E}_{B]}{}^N \qquad (4.19)$$

especially for later embedding in the SL(5) of F-theory.

The coordinate transformations are then

$$\delta \mathcal{E}_A{}^M = \tfrac{1}{2}\lambda^{NP}\partial_{NP}\mathcal{E}_A{}^M - \mathcal{E}_A{}^N(\partial_{NP}\lambda^{MP} - \tfrac{1}{4}\delta^M_N \partial_{PQ}\lambda^{PQ}) \qquad (4.20a)$$

$$\delta \Phi = \tfrac{1}{2}\lambda^{MN}\partial_{MN}\Phi + \tfrac{1}{4}\Phi \partial_{MN}\lambda^{MN} \qquad (4.20b)$$

$$\delta R_M = \tfrac{1}{2}\lambda^{NP}\partial_{NP}R_M + (\partial_{MP}\lambda^{NP})R_N \qquad (4.20c)$$

The non-transport term on $\mathcal{E}$ is an SL(4) transformation. (Note that for $R$ we have $\partial_{MP}\lambda^{NP} = (\partial_{MP}\lambda^{NP} - \tfrac{1}{4}\delta^N_M \partial_{PQ}\lambda^{PQ}) + \tfrac{1}{4}\delta^N_M \partial_{PQ}\lambda^{PQ}$.) We also have the R-R gauge transformations

$$\delta R_M = \partial_{MN}\Lambda^N \qquad (4.21)$$

We can again derive the $R$ coordinate transformations from the gauge covariant version,

$$\delta R_M = \tfrac{1}{2}\lambda^{NP}F_{MNP} \,, \quad F_{MNP} = \tfrac{1}{2}\partial_{[MN}R_{P]} \qquad (4.22)$$

or similar expressions with $F^M = \frac{1}{6}\varepsilon^{MNPQ} F_{NPQ}$.

It's convenient to decompose indices as

$$V_{[AB]} = (V_{1a}, V_{[ab]}) \equiv (V_a, \varepsilon_{abc} V^c) \to (V_a, V^a) \tag{4.23}$$

and similarly for curved indices, where $A = (1, a)$, $a = (2, 3, 4)$. This gives the desired form of the SO(3,3) metric

$$V^2 = \tfrac{1}{4}\varepsilon^{ABCD} V_{AB} V_{CD} \to 2 V^a V_a \tag{4.24}$$

This allows us to straightforwardly identify components of $\mathcal{E}$ with those of $E$, and we find

$$\mathcal{E}_A{}^M = \frac{1}{a}\begin{pmatrix} e^{1/2} & e^{1/2} \tfrac{1}{2}\varepsilon^{mnp} B_{np} \\ -e^{-1/2} e_a{}^m \tfrac{1}{2}\varepsilon_{mnp} B^{np} & e^{-1/2} e_a{}^m \end{pmatrix} \tag{4.25a}$$

$$= \begin{pmatrix} e^{1/2} & 0 \\ 0 & e^{-1/2} e_a{}^n \end{pmatrix} \begin{pmatrix} 1 & \tfrac{1}{2}\varepsilon^{mpq} B_{pq} \\ -\tfrac{1}{2}\varepsilon_{npq} B^{pq} & \delta_n^m \end{pmatrix} \tag{4.25b}$$

with a gauge condition again breaking $SO(2,1)^2$ to $SO(2,1)$.

Under this decomposition, the R-R gauge transformations branch into

$$\delta R_m = \partial_m \Lambda + \varepsilon_{mnp} \partial^n \Lambda^p, \quad \delta R = \partial_m \Lambda^m \tag{4.26}$$

Depending on how we section to S-theory (which $X$'s survive, $x^m$ or $x_m$), this identifies these fields as either

$$\partial_m: \quad R_m = \text{1-form}, \quad R = \text{3-form} \quad \text{(IIA)} \tag{4.27a}$$
$$\partial^m: \quad R_m = \text{2-form}, \quad R = \text{0-form} \quad \text{(IIB)} \tag{4.27b}$$

This illustrates that both theories arise from different choices of section of the same theory (and the fact that T-duality relates them).

To avoid redefining the NS-NS fields (and switching up and down indices), one can also start with different chirality spinors, and section the same way:

$$R_M \Rightarrow R_m = \text{1-form}, \quad R = \text{3-form} \quad \text{(IIA)} \tag{4.28a}$$
$$R^M \Rightarrow R^m = \text{2-form}, \quad R = \text{0-form} \quad \text{(IIB)} \tag{4.28b}$$

(with corresponding changes in the gauge transformations from $R_M$ to $R^M$). But F-theory doesn't prefer this.

A third alternative is to introduce both $R_M$ and $R^M$, and relate them by duality. Since this is enforced via field equations, it requires introduction of the bosonic coset: In a flat background, the condition would break SO(D,D) to SO(D−1,1)$^2$ (just as selfduality of vectors in 4D N = 8 supergravity requires the bosons to avoid breaking E$_7$ to SU(8) and GL(4) to SO(3,1)).

Although SO(2,1) spinors aren't chiral, SO(3,3) are, so in this way there can be 2 sections of Type II T-theory to S-theory.

## 4.3 Type I

We now impose also parity projection on both G = SO(D,D) and H = SO(D−1,1)$_L$×SO(D−1,1)$_R$, where the parity matrix $\mathcal{P}$ in both bases is

$$\mathcal{P}_A{}^B = \begin{pmatrix} 0 & \delta_a^{\bar{b}} \\ \delta_{\bar{a}}^{b} & 0 \end{pmatrix}, \quad \mathcal{P}_M{}^N = \begin{pmatrix} \delta_m^n & 0 \\ 0 & -\delta_n^m \end{pmatrix} \tag{4.29}$$

(clearly switching L and R, and changing $\sigma \to -\sigma$), and parity projection on $E$ and an element of the H gauge group $\Lambda$ is

$$\mathcal{P}_A{}^B E_B{}^N \mathcal{P}_N{}^M = E_A{}^M, \quad \mathcal{P}_A{}^C \Lambda_C{}^D \mathcal{P}_D{}^B = \Lambda_A{}^B \tag{4.30}$$

(It's actually easier to solve $\mathcal{P}$ projection before orthogonality.) The combination of these 2 conditions, besides identifying the L and R parts of H, also constrains

$$E_A{}^M = \frac{1}{\sqrt{2}} \begin{pmatrix} e_a{}^m & -e_{ma} \\ e_a{}^m & e_{ma} \end{pmatrix} \tag{4.31}$$

in the mixed basis or

$$E_A{}^M = \begin{pmatrix} e_a{}^m & 0 \\ 0 & e_m{}^a \end{pmatrix} \tag{4.32}$$

in the co/contra-variant basis for both. So the coset is reduced to GL(D)/SO(D−1,1) of ordinary gravity.

Then

$$\eta M \eta = M^{-1}, \quad \mathcal{P} M \mathcal{P} = M \quad \Rightarrow \quad M^{MN} = \begin{pmatrix} G^{mn} & 0 \\ 0 & G_{mn} \end{pmatrix} \tag{4.33}$$

Parity projection on $R$ in components (or the Ramond-Neveu-Schwarz formalism) is not so transparent: On field strengths, one simply antisymmetrizes on left and right spinor indices. (NS-NS fields are symmetrized in vector indices, R-R are antisymmetrized in spinor indices, according to graded symmetrization.) On the other hand the physical, transverse R-R degrees of freedom are given by antisymmetrizing spinor indices of the little group. However, since F-theory unifies NS-NS and R-R fields into a single gauge field, the procedure there is straightforward, as we'll see below, and the decomposition of F Type I to T Type I will illustrate the R-R projection. The reason for the difference is that branes, unlike the string worldsheet, can carry R-R charges, so background fields must include the gauge fields explicitly. (In [23] extra currents for R-R charges were introduced into T-theory. These currents are automatically incorporated in F-theory, as part of the same exceptional symmetry representation that includes the currents for the NS-NS "charges": All bosonic currents are "unified".)

Note that in D = 3, the R-R field strength in Type I is given by antisymmetrizing 2-valued spinor indices of SO(2,1), and thus a scalar, which is the same as a 3-form, the field strength of a 2-form. On the other hand, the physical R-R degrees of freedom are given by antisymmetrizing 1-valued spinor indices of SO(1), which gives nothing, because a 2-form has none.

## 5 3D F-theory

### 5.1 Type II

For F-theory we again look at the simple case of D = 3, which is actually simpler than the general T-theory case.

The solution to the generalized orthogonality condition [9]

$$E_{AB}{}^{MN} = \mathcal{E}_{[A}{}^M \mathcal{E}_{B]}{}^N \tag{5.1}$$

is simply the statement that the (inverse) worldvolume vielbein $\mathcal{E}$ is just a different SL(5)/SO(3,2) representation of (the bosonic part of) the spacetime vielbein $E$ ($\partial^M = \partial/\partial \sigma_M$):

$$\triangleright_{AB} = \tfrac{1}{2} E_{AB}{}^{MN} \mathring{\triangleright}_{MN}, \quad \mathcal{D}^A = \mathcal{E}_M{}^A \partial^M \tag{5.2}$$

The gauge transformation is

$$\delta \mathcal{E}_A{}^M = \tfrac{1}{2} \lambda^{NP} \partial_{NP} \mathcal{E}_A{}^M - \mathcal{E}_A{}^N (\partial_{NP} \lambda^{MP} - \tfrac{1}{5} \delta_N^M \partial_{PQ} \lambda^{PQ}) \tag{5.3}$$

The non-transport term is SL(5).

The goal is to reduce Type II F-gravity to Type I and heterotic T-gravities. The procedure we described was:

(1) First parity project Type II F-theory with respect to all $\sigma$ but $\sigma_2$ to Type I F-theory.

(2) ($\mathcal{U}$-)Section to preserve $\sigma_1$ as the surviving $\sigma$ for Type I T-theory and $\sigma_2$ for heterotic.

However, the simplest way to find these theories is to instead

(a) First apply $\sigma_2$-preserving sectioning to Type II F-gravity, producing the known Type II T-gravity (NS-NS and R-R sectors).

(b) Then parity project by reflecting all but $\sigma_2$ labels for heterotic or $\sigma_1$ labels for Type I. (This is a reversal of the labeling of $\sigma_1$ and $\sigma_2$ for the Type I case.)

$$
\begin{array}{ccc}
 & \text{FII} & \\
\mathcal{P} \swarrow & & \searrow \mathcal{U} \\
\text{FI} & & \text{TII} \\
\searrow & & \swarrow \\
 & \text{TI, TH} &
\end{array}
\tag{5.4}
$$

In this subsection, we apply the first step, finding Type II T-gravity. For D = 3, this means separating the indices as $A \to (2, a)$ ($a = 1, 3, 4, 5$), thus separating $\mathcal{E}$ into NS-NS, R-R, and dilaton fields, as the SL(4)/SO(2,2) fields described above.

We then drop all $X^{[MN]}$'s with $M$ or $N = 2$. Looking at the F-gravity coordinate transformation, for the moment we can ignore the transport term $\lambda \bullet \partial$ and density term $\partial \bullet \lambda$. We then see that dropping $\partial_{2m}$ implies that fixing the Lorentz gauge $\mathcal{E}_2{}^m = 0$ is preserved by the coordinate transformation, while breaking SO(3,2) $\to$ SO(2,2). We next identify the dilaton $\Phi$ as the only scalar (density, with the appropriate weight), $\mathcal{E}_a{}^m \in$ SL(4)/SO(2,2) in the defining representation, and the R-R field $R_m$ as a spinor, all transforming as expected for Type II T-gravity: As above,

$$\delta \mathcal{E}_a{}^m = \tfrac{1}{2}\lambda^{np}\partial_{np}\mathcal{E}_a{}^m - \mathcal{E}_a{}^n(\partial_{np}\lambda^{mp} - \tfrac{1}{4}\delta_n^m \partial_{pq}\lambda^{pq}) \tag{5.5a}$$

$$\delta \Phi = \tfrac{1}{2}\lambda^{mn}\partial_{mn}\Phi + \tfrac{1}{4}\Phi \partial_{mn}\lambda^{mn} \tag{5.5b}$$

$$\delta R_m = \tfrac{1}{2}\lambda^{np}\partial_{np}R_m + (\partial_{mp}\lambda^{np})R_n + \partial_{mn}\lambda^{n2} \tag{5.5c}$$

The result is

$$
\mathcal{E}_A{}^M = \begin{array}{c} 2 \\ a \end{array}\begin{pmatrix} \overset{2}{\Phi^{4/5}} & \overset{m}{0} \\ \Phi^{-1/5}\mathcal{E}_a{}^m R_m & \Phi^{-1/5}\mathcal{E}_a{}^m \end{pmatrix} = \begin{pmatrix} \Phi^{4/5} & 0 \\ 0 & \Phi^{-1/5}\mathcal{E}_a{}^n \end{pmatrix}\begin{pmatrix} 1 & 0 \\ R_n & \delta_n^m \end{pmatrix}
\tag{5.6}
$$

Then

$$
E_{AB}{}^{MN} = \mathcal{E}_{[A}{}^M \mathcal{E}_{B]}{}^N = \tfrac{2a}{ab}\begin{pmatrix} \overset{2m}{\Phi^{3/5}\mathcal{E}_a{}^m} & \overset{mn}{0} \\ \Phi^{-2/5}E_{ab}{}^{mn}R_n & \Phi^{-2/5}E_{ab}{}^{mn} \end{pmatrix}
\tag{5.7}
$$

where $E_{ab}{}^{mn} = \mathcal{E}_{[a}{}^{m}\mathcal{E}_{b]}{}^{n} \in$ SO(3,3)/SO(2,1)$^2$ in its defining representation. $\mathcal{E}$ is more convenient for algebra, but $E$ relates more to general D. In particular, a similar decomposition was found in [23], where currents coupling to R-R charges were added to 10D T-theory.

This $E_{ab}{}^{mn}$ appears as the same type of coset in Type I F-theory, but $X$ and $\sigma$ are bigger.

## 5.2 Type I and heterotic

To be slightly more general (for later applications), we examine the coset GL(N)/SO(N) (without regard to signature), and consider a parity transformation that changes an arbitrary number of signs n,

$$\mathcal{P} = \begin{pmatrix} I & 0 \\ 0 & -I \end{pmatrix} \tag{5.8}$$

Then

$$\mathcal{P}\begin{pmatrix} a & b \\ c & d \end{pmatrix}\mathcal{P} = \begin{pmatrix} a & b \\ c & d \end{pmatrix} \quad \Rightarrow \quad b = c = 0 \tag{5.9}$$

simply forces the matrix to be block diagonal. (This is also clear from the index structure.) Thus

$$\frac{\text{GL(N)}}{\text{SO(N)}} \quad \to \quad \frac{\text{GL(N$-$n)GL(n)}}{\text{SO(N$-$n)SO(n)}} \tag{5.10}$$

In the present case, this means

$$\frac{\text{SL}(5)}{\text{SO}(3,2)} \quad \to \quad \frac{\text{SL}(4)}{\text{SO}(2,2)}\text{GL}(1) \tag{5.11}$$

Applying this restriction to the above Type II expressions,

$$\mathcal{E}_A{}^M = \begin{array}{c} 2 \\ a \end{array}\begin{pmatrix} \overset{2}{\Phi^{4/5}} & \overset{m}{0} \\ 0 & \Phi^{-1/5}\mathcal{E}_a{}^m \end{pmatrix} \tag{5.12}$$

and

$$E_{AB}{}^{MN} = \mathcal{E}_{[A}{}^M\mathcal{E}_{B]}{}^N = \begin{array}{c} 2a \\ ab \end{array}\begin{pmatrix} \overset{2m}{\Phi^{3/5}\mathcal{E}_a{}^m} & \overset{mn}{0} \\ 0 & \Phi^{-2/5}E_{ab}{}^{mn} \end{pmatrix} \tag{5.13}$$

Since the heterotic coset is the same as Type II T-theory (NS-NS sector), identifying $\sigma_2$ as the surviving $\sigma$ produces a T-theory action identical to that for the NS-NS sector of Type II [8], as expected, since NS-NS is the entire bosonic sector of heterotic string supergravity (neglecting the Yang-Mills supermultiplets).

To compare Type I T-theory to heterotic, we perform the same decomposition as for Type II T-theory, distinguishing the "1" label from the remaining. Combining the above results for F II → T II and T II → ordinary fields, (now $m = 3, 4, 5$)

$$\mathcal{E}_A{}^M = \begin{array}{c} 2 \\ 1 \\ a \end{array} \begin{pmatrix} \overset{2}{\Phi} & \overset{1}{0} & \overset{m}{0} \\ e^{1/2}(R_1 + \tfrac{1}{2}\varepsilon^{mnp}B_{mn}R_p) & e^{1/2} & e^{1/2}\tfrac{1}{2}\varepsilon^{mnp}B_{np} \\ e^{-1/2}e_a{}^m(R_m - \tfrac{1}{2}\varepsilon_{mnp}B^{np}R_1) & -e^{-1/2}e_a{}^m\tfrac{1}{2}\varepsilon_{mnp}B^{np} & e^{-1/2}e_a{}^m \end{pmatrix} \Phi^{-1/5}$$

(5.14)

with $(R_1, R_m)$ as either odd (1 and 3) forms for IIA ($B^{mn} = 0$) or even (0 and 2) for IIB ($B_{mn} = 0$) upon reduction to S-theory:

$$R_1 + \tfrac{1}{2}\varepsilon^{mnp}B_{mn}R_p = \begin{cases} \varepsilon^{mnp}(\tfrac{1}{6}C_{mnp} + \tfrac{1}{2}B_{mn}A_p) & \text{(IIA)} \\ \tilde{\varphi} & \text{(IIB)} \end{cases} \quad (5.15a)$$

$$R_m - \tfrac{1}{2}\varepsilon_{mnp}B^{np}R_1 = \begin{cases} A_m & \text{(IIA)} \\ \tfrac{1}{2}\varepsilon_{mnp}(\tilde{B}^{np} - B^{np}\tilde{\varphi}) & \text{(IIB)} \end{cases} \quad (5.15b)$$

We then drop all components of $\mathcal{E}_A{}^M$ with an odd number of $A$ and $M = 2$ for heterotic or 1 for Type I. Thus both N = 1 theories have $e_a{}^m$ and $\Phi$, but heterotic has $B_{mn}$ while Type I has $\tilde{B}^{mn}$. As we already saw in Type II T-theory, coordinate transformations required that $B$ (NS-NS) couples to the dilaton $\Phi$, while $\tilde{B}$ (R-R) doesn't. (But the dilaton doesn't appear as an independent field in F-theory or M-theory, as is the case in 11D supergravity.)

Parity projection is with respect to both indices and coordinates: This picks out IIB rather than IIA, which we have distinguished in the 3D case with respect to coordinate dependence.

The fact that Type II supergravity can be projected in 2 different ways to get either Type I or heterotic supergravity is not new. What is new is that Type II F-theory can be projected to get Type I F-theory, which has 2 different sections that describe either Type I or heterotic string theory.

## 5.3 Stringy supergravities

Besides being simpler than their 10D analogs, in 3D (1) N = 2 is nonmaximal, so the supermultiplet is reducible, and (2) auxiliary fields and superspace formulations are known.

The Type IIA theory is known from M-theory (4D N=1 supergravity with a 3-form gauge auxiliary) and T-theory [34], F-theory [35], and our rederivation above: N = 2 supergravity coupled to a vector-tensor multiplet. The dimension-0 field content is the graviton and differential forms,

$$e_a{}^m, \quad \varphi, \quad A_m, \quad B_{mn}, \quad C_{mnp} \quad (5.16)$$

(The fermions are the usual gravitini and spin-½. The dimension-1 non-gauge auxiliary fields are a vector and scalars.) This separates into fields for pure supergravity

$$e_a{}^m, \quad C_{mnp} \tag{5.17}$$

and the matter multiplet

$$\varphi, \quad A_m, \quad B_{mn} \tag{5.18}$$

The reduction of N = 2 supergravity to N = 1 is unique (drop $C$), but the vector-tensor multiplet breaks into an N = 1 vector multiplet

$$A_m \tag{5.19}$$

and an N = 1 tensor multiplet

$$\varphi, \quad B_{mn} \tag{5.20}$$

By the usual construction of heterotic supergravity as vector × vector multiplet without symmetrization, its content is

$$e_a{}^m, \quad \varphi, \quad B_{mn} \tag{5.21}$$

using the tensor multiplet.

On the other hand (chirality) Type IIB, as described above, uses for the N = 2 matter a variant of the scalar multiplet: The total dimension-0 field content

$$e_a{}^m, \quad \varphi, \quad \widetilde{\varphi}, \quad B_{mn}, \quad \widetilde{B}_{mn} \tag{5.22}$$

divides up into that for N = 1 supergravity

$$e_a{}^m \tag{5.23}$$

and for matter

$$\varphi, \quad \widetilde{\varphi}, \quad B_{mn}, \quad \widetilde{B}_{mn} \tag{5.24}$$

That latter branches into 2 N = 1 tensor multiplets

$$\varphi, \quad B_{mn}; \quad \widetilde{\varphi}, \quad \widetilde{B}_{mn} \tag{5.25}$$

As we saw above, Type I then chooses 1 of these multiplets, using the same scalar (dilaton) as heterotic, but a different 2-form.

In summary, we have from Type II theories to Type I theories in the D = 3 case:

| | st | wv | coset+ | fields | $\mathcal{P}$(arity) | coset+ | fields |
|---|---|---|---|---|---|---|---|
| | | | | Type II | | Type I | |
| F | $X^{MN}$ $M$=1-5 $10$ | $\sigma_M, \tau$ $M$=1-5 $5+1$ | $\frac{SL(5)}{SO(3,2)}$ | $\mathcal{E}_A{}^M$ $A,M$=1-5 $14$ | $\sigma_2 \to \sigma_2$ $\sigma_m \to -\sigma_m$ $m$=1,3,4,5 | $\frac{SL(4)}{SO(2,2)}$ $+1$ | $\mathcal{E}_a{}^m, \Phi$ $a,m$=1,3,4,5 $9+1$ |
| M | $X^M$ $M$=1-4 $4$ | $\sigma_M, \tau$ $M$=1-4 $4+1$ | $\frac{GL(4)}{SO(3,1)}$ $+4$ | $e_A{}^M, A_{MNP}$ $A,M$=1-4 $10+4$ | $\sigma_2 \to \sigma_2$ $\sigma_m \to -\sigma_m$ $m$=1,3,4 | $\frac{GL(3)}{SO(2,1)}$ $+1+3$ | $e_a{}^m, \varphi, B_{mn}$ $a,m$=1,3,4 $6+1+3$ |
| T | $X^{MN}$ $M$=1-4 $6$ | $\sigma, \tau$ $1+1$ | $\frac{O(3,3)}{SO(2,1)^2}$ $+1+4$ | $E_{AB}{}^{MN}, \Phi, R_M$ $A,M$=1-4 $9+1+4$ | $\sigma \to -\sigma$ | $\frac{GL(3)}{SO(2,1)}$ $+1+3$ | $e_a{}^m, \varphi, R_m$ $a,m$=1,2,3 $6+1+3$ |

(5.26)

All the different theories can now be represented as

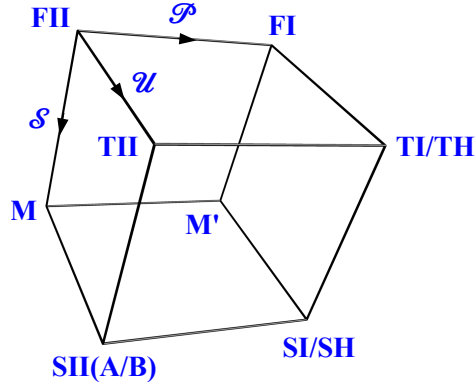

For clarity, we haven't split some theories, and therefore didn't indicate that (1) M → SIIA only, (2) M′ → SH only, and (3) only SIIB → SI/SH. We can also just list the theories by which combinations of $\mathcal{P}, \mathcal{S}, \mathcal{U}$ have been applied:

$$\begin{array}{ll} - & \text{FII} \\ \mathcal{P} & \text{FI} \\ \mathcal{S} & \text{M} \\ \mathcal{U} & \text{TII} \\ \mathcal{P}, \mathcal{S} & \text{M}' \\ \mathcal{P}, \mathcal{U} & \text{TI/TH} \\ \mathcal{S}, \mathcal{U} & \text{SII(A/B)} \\ \mathcal{P}, \mathcal{S}, \mathcal{U} & \text{SI/SH} \end{array}$$

(5.27)

# 6 Conclusions

There are a number of topics for further study:

(1) We have assumed a restricted topology. For $\sigma_1$ and $\sigma_2$, we considered only cylinders (open branes) and donuts (closed branes), which are the only orientable surfaces with vanishing Euler number. This implies some kind of constraint on curvature. We also assumed tori with respect to the remaining worldvolume dimensions, but did not make an explicit application.

(2) There are some results that will probably require quantization and consideration of anomalies: e.g., opposite chirality of $\Theta$ and internal degrees of freedom, and choice of Yang-Mills symmetry groups. (Previously we have quantized only indirectly by first reducing to S-theory.)

(3) In particular, what precludes the existence of $E_8 \times E_8$ Type I superstrings, since the corresponding open F-brane is apparently the origin of the corresponding heterotic superstring?

(4) Alternatively, might the Yang-Mills symmetry groups follow as solutions of the quantum corrected background field equations for the metric of the 496 scalars? Quantum F-theory isn't yet understood (except on sectioning to S-theory), but maybe some semiclassical arguments would suffice. Alternatively, the usual determination by anomalies would require only the massless sector of F-theory.

(5) What are symmetry relations or dualities between Type I and Type II branes? Can symmetry breaking be attributed to super-Higgs? What's the relation to the boundary currents?

(6) What are symmetry relations or dualities between open and closed Type I branes? In particular, how can supergravity come from not only closed branes (Type I superstrings), but also open (heterotic superstrings)?

In future papers we plan to address these questions, as well as give details on

(a) the worldvolume formulation,

(b) the F-gravity action,

(c) the D = 4 case, and

(d) seeing if the reduction of the symmetry groups from exceptional groups to SO(D,D) allows a simple treatment of the general case.

# Acknowledgements

WS is supported by NSF grant PHY-1915093. We thank Dharmesh Jain for help with the bibliography, and his style type `hephys`.

# A Cosets

The cosets F/L for the Lagrangians and G/H for the Hamiltonians are [14]:

```
    F
τ ↙   ↘ |0⟩
G         L
  ↘   ↙
    H
```

| D | d | F | G | L | H | |
|---|---|---|---|---|---|---|
| 0 | 2 | GL(2) | GL(1) | GL(1,**C**) | I | |
| 1 | 3 | GL(3) | GL(2) | GL(2) | SO(1,1) | |
| 2 | 4 | SL(4)SL(2) | SL(3)SL(2) | GL(2)$^2$ | GL(2) | |
| 3 | 6 | SL(6) | SL(5) | GL(4) | Sp(4) | (A.1) |
| 4 | 12 | SO(6,6) | SO(5,5) | GL(4,**C**) | Sp(4,**C**) | |
| 5 | 56 | $E_{7(7)}$ | $E_{6(6)}$ | U*(8) | USp(4,4) | |
| 6 | ? | ? | $E_{7(7)}$ | U*(8)$^2$ | SU*(8) | |
| 7 | ? | ? | $E_{8(8)}$ | U*(16) | SO*(16) | |

(Reduction from Lagrangian to Hamiltonian is indicated by $\tau$, spontaneous symmetry breaking by the vacuum $|0\rangle$.) The question marks are for infinite-dimensional extended Lie algebras, as indicated by the Dynkin diagram in the main text.

# B  F → T and M

We present here a counting comparison of the fields in Type II F-theory, from G/H, and T-theory, from SO(D,D)/SO(D−1,1)$^2$ = D$^2$ (NS-NS) + SO(D,D) Weyl spinor = $2^{D-1}$ (R-R). The discrepancy is attributed to the dilaton (compensating) multiplet, which comes from the ghost-ghost sector. We also include M-theory, as vielbein GL(D+1)/SO(D,1) + 3-form + 6-form in 1 extra dimension.

| D | F | T | M | |
|---|---|---|---|---|
| 0 | 1 − 0 = 1 | 0 + 0 = 0 | 1 + 0 = 1 | |
| 1 | 4 − 1 = 3 | 1 + 1 = 2 | 3 + 0 = 3 | |
| 2 | 11 − 4 = 7 | 4 + 2 = 6 | 6 + 1 = 7 | |
| 3 | 24 − 10 = 14 | 9 + 4 = 13 | 10 + 4 = 14 | (B.1) |
| 4 | 45 − 20 = 25 | 16 + 8 = 24 | 15 + 10 = 25 | |
| 5 | 78 − 36 = 42 | 25 + 16 = 41 | 21 + 20 + 1 = 42 | |
| 6 | 133 − 63 = 70 | 36 + 32 = 68 | 28 + 35 + 7 = 70 | |
| 7 | 248 − 120 = 128 | 49 + 64 = 113 | 36 + 56 + 28 = 120 | |

Note that the M-theory counting agrees with F except for D = 7 (lacking an SO(7,1) vector), and T lacks only the dilaton for D < 6 (D = 6 needs an extra scalar, D = 7 an extra SO(7,7) vector).